\documentclass[a4paper]{article}
\usepackage[font=small]{caption}
\usepackage{bbm}
\usepackage[english]{babel}
\usepackage[utf8]{inputenc}
\usepackage{amsthm}
\usepackage{graphicx}
\usepackage{subfigure}
\usepackage{float}
\usepackage{epsfig}
\usepackage[colorinlistoftodos]{todonotes}
\usepackage[hidelinks]{hyperref}
\usepackage[margin=0.8in]{geometry}
\usepackage{amssymb}
\usepackage[ruled,vlined,linesnumbered]{algorithm2e}
\usepackage{babel}
\usepackage{amsmath}
\usepackage{enumerate}
\usepackage{xcolor}
\usepackage{optidef}

\usepackage{authblk}
\usepackage{natbib}

\usepackage{float}
\usepackage{color}
\usepackage{diagbox}
\usepackage{comment}

\usepackage{threeparttable}

\newtheorem{proposition}{Proposition}

\usepackage{blindtext}

\makeatletter
\renewcommand{\paragraph}{\@startsection{paragraph}{4}{0ex}%
    {-3.25ex plus -1ex minus -0.2ex}%
    {1.5ex plus 0.2ex}%
    {\normalfont\normalsize\bfseries}}
\makeatother

\stepcounter{secnumdepth}
\stepcounter{tocdepth}

\begin{document}

\title{Optimal market completion through financial derivatives with applications to volatility risk. }

\author[1,2]{Matt Davison\thanks{Email address: mdavison@uwo.ca}}
\author[1]{Marcos Escobar-Anel\thanks{Email address:  marcos.escobar@uwo.ca}}
\author[1]{Yichen Zhu\thanks{Email address: yzhu562@uwo.ca}}

\affil[1]{Department of Statistical and Actuarial Sciences, \ Western University}
\affil[2]{Department of Mathematics, \ Western University}

\date{\today}

\maketitle
\begin{abstract}
This paper investigates the optimal choices of financial derivatives to complete a financial market in the framework of stochastic volatility (SV) models. We introduce an efficient and accurate simulation-based method, applicable to generalized diffusion models, to approximate the optimal derivatives-based portfolio strategy. We build upon the double optimization approach (i.e. expected utility maximization and risk exposure minimization) proposed in \cite{2022arXiv220103717E}; demonstrating that strangle options are the best choices for market completion within equity options. Furthermore, we explore the benefit of using volatility index derivatives and conclude that they could be more convenient substitutes when only long-term maturity equity options are available.
\end{abstract}

\section{Introduction}

Financial markets are often modelled as a system of contingents on states mirroring the real-world economy. This generates a concept widely used in economic and finance literature, namely the complete market, which is simply described as `a market for every good'. Earlier studies assumed that the number of securities equals the number of states of nature and investigated the optimal allocation, placing all of the capital at once (see \cite{arrow1964role}, \cite{arrow1954existence}). Recognizing that investors benefit from adjusting allocation with a change of market status, more recent researchers have focused on the idea of a dynamically complete market, which is defined as a market wherein any contingent claim can be replicated by a self-financing strategy. 

The study of portfolio choice in a dynamically complete market under a continuous-time framework can be traced back to the seminal work of \cite{merton1969lifetime}, who computed the optimal allocation and consumption policy with a dynamic programming technique, assuming that the stock price follows a geometric Brownian motion (GBM). In this framework, the uncertainty is reflected in the Brownian motion, which captures the randomness of a stock’s return; hence, investors can achieve the best portfolio performance with investments only on the stock and a cash account. 

However, the financial market is ever evolving, and becoming increasingly complex; for instance, substantial evidence suggests that a single Brownian motion or source of randomness is insufficient to explain the movements of a single stock or index. Researchers have had to incorporate so-called stylized facts such as stochastic volatility (SV) or stochastic interest rates in their modelling to mimic this new reality. These stylized facts are captured via adding new `state variables' (e.g. a new random processes for SV). These state variables have been recognized as important factors in the portfolio allocation process. 

The importance of adding financial derivatives into a portfolio for market completion was demonstrated in \cite{liu2003dynamic}, confirming that investors can improve portfolio performance when adding as many linearly independent equity options as new state variables in the portfolio composition. They do this to hedge the risk of the new state variables, thereby achieving significant improvement in portfolio performance compared to incomplete market investment (e.g. investing solely on stock and cash account). This work was extended in many directions. For example, \cite{escobar2017optimal} constructed an optimal portfolios with the addition of options to hedge new state variables accounting for stochastic correlation. Moreover, \cite{li2018dynamic} solved derivative-based strategies under an asset–liability management (ALM) framework with the mean–variance criterion. In a similar setting, the optimal complete and incomplete strategies for the 4/2 SV model were derived in \cite{cheng2021optimal}, which demonstrated the superiority of the complete market portfolio.  

Although the literature cited above strongly supports the addition of derivatives to complete the market, investors may complete market in many ways due to the variety of derivatives in the market. Therefore, investors effectively have a non-unique solution to the problem (i.e. an infinite number of strategies, each linked to a choice of derivative, producing the same maximum expected utility). The problem of infinitely many solutions and the optimal choice of derivatives was studied in the recent paper \cite{2022arXiv220103717E} in the context of the Black---Scholes---Merton model. The paper proposed an  optimization criterion (i.e. additional to the maximization of the utility, namely risk exposure minimization) to produce a unique, meaningful solution, thus deriving a practical derivative selection methodology for investors. The risk exposure minimization criterion can be motivated from many angles, especially in terms of regulatory constraints intended to control investors' exposure to risky assets and hence to protect investors’ capital in the event of a market crash. In this paper, we follow the same derivatives selection framework and explore the optimal product for market completion in the popular setting of SV models, with emphasis on the celebrated Heston model (see \cite{heston1993closed}).

There are two major hurdles for our derivatives-based portfolio allocation problem. First, given the complexity of advanced models with many state variables jeopardizes the solvability of the utility maximization allocation problem, closed-form solutions are often unavailable. This hurdle can be overcome using approximation methods for dynamic portfolio choice problems. \cite{brandt2005simulation}, inspired by the least-squares Monte Carlo method (see \cite{longstaff2001valuing}), recursively estimated the value function and optimal allocation following a dynamic programming principle. This method was later named the BGSS and \cite{cong2017accurate} utilized the stochastic grid bundling method for conditional expectation estimation, introduced in \cite{jain2015stochastic}, further enhancing the accuracy of BGSS. Additionally, \cite{zhu2022polynomial} targeted unsolvable continuous-time models, proposing an efficient and accurate simulation-based method, namely the polynomial affine method for constant relative risk aversion utility (PAMC). The second hurdle appears in the complexity of derivatives’ price dynamics which, in contrast to traditional asset classes, could lead to highly non-linear stochastic differential equations. In this paper, we overcome the two hurdles simultaneously by unifying the PAMC and an options’ Greek approximation technique. Notably, the broad applicability of this methodology laid the foundation for the derivatives selection study within a generalized model family.

As mentioned above, our focus is on investors who are particularly concerned about volatility risk and seek the best derivatives to attain market completion. The seminal paper by \cite{heston1993closed} recognized the mean-reverting pattern of volatilities and introduced the well-known Heston (GBM 1/2) model. Later, extensions, such as the GBM 3/2 (see \cite{heston1997simple}) and GBM 4/2 (see \cite{grasselli20174}), were developed to better capture the volatility surface. These lead to notable successes in the valuation of European equity options, and semi-closed-form solutions for the option price and Greeks are generally accessible using Fourier transformation. Popular equity options, such as call, put, straddle and strangle options, are ideal products for investors to manage the volatility risk. Furthermore, the volatility index (VIX), a measure of the stock market's  volatility based on S\&P 500 index options provided by the Chicago Board Options Exchange (CBOE),  affords investors an alternative way to assess the volatility risk. The effectiveness of VIX products in the portfolio performance enhancement has been confirmed in the literature: see \cite{doran2020volatility}, \cite{chen2011diversification} and \cite{warren2012can}. Hence, in this paper we compare two categories of derivatives, namely equity options and VIX options in terms of optimal dynamic completion. 

The contributions of the paper are as follows:
\begin{enumerate}
    \item The multitude of financial derivatives available in the market offers investors non-unique optimal choice in terms of expected utility theory (EUT) maximization. Hence we extend the additional optimization criterion proposed in \cite{2022arXiv220103717E}, namely risk exposure minimization, from the family of GBM to SV models. This aids investors with practical derivative selection in a popular stock markets modeling setting.
    \item The PAMC-indirect numerical method is proposed to approximate the optimal allocation for a constant relative risk aversion (CRRA) investor investing in the derivatives market. The superior accuracy and efficiency of the methodology are verified on the Heston model. 
    \item Targeting equity and volatility risk, we first consider the optimal choice among equity options (e.g. calls, puts, straddles and strangles). We demonstrate that strangles are the best options for minimizing risk exposure.   
    \item  We also investigate the usage of financial derivatives on the VIX as a means of completing the market, and we conclude that investors would prefer VIX options to equity strangles when only long-term maturity options are available. 

\end{enumerate}
The remainder of this paper is organized as follows: Section \ref{chp_6_sec2} presents the investor's problem (i.e. the two criteria for optimal allocation [utility maximization] and optimal market completion [risk exposure minimization]).  Section \ref{chp_6_sec3} details an efficient approximation method for derivatives-based portfolio allocation. The optimal market completion targeting volatility risk within an equity option and a VIX option is studied in Section \ref{chp_6_sec4}, followed by the conclusion in Section \ref{chp_6_sec5}. Appendix \ref{chp6_appendix_proof} presents the mathematical proofs, while Appendix \ref{chp6_direct} provides an alternative approximation method and a numerical examination of accuracy and  efficiency for the two methods.

\section{Investor's problem}
\label{chp_6_sec2}
In this section, we introduce a market completion framework using financial derivatives. We define a complete probability space $(\Omega ,\mathcal{F},\mathbb{P})$ with
a right-continuous filtration $\{\mathcal{F}_{t}\}_{t\in \lbrack 0,T]}$. The market is frictionless (i.e. no transaction cost and market impact), and a risk-free cash account $M_t$, a stock $S_t$ and an investor with constant relative risk aversion (CRRA) utility, $U(W)=\frac{W^{1-\gamma}}{1-\gamma}$ exist. The market dynamics are summarized as follows:
\begin{equation}
\begin{cases}
\frac{d M_t}{M_t}=rdt\\
\frac{d S_t}{S_t}=(r+\lambda^S\sigma^S )dt+\sigma^S  dB_t^S\\
d H_t=\mu^Hdt+\sigma^H dB_t^H \\
<dB_t^S,dB^H_t>=\rho_{SH} dt.
\end{cases}\label{chp6_gene_diff_model}
\end{equation}
 where $B_t^H$ and $B_t^S$ are Brownian motions with correlation $\rho_{SH}\in (-1,1)$, and the interest rate $r$ is constant. State variable $H_t$ follows a generalized diffusion process, where $\mu^H=\mu^H(t,H_t)$ denotes the drift and $\sigma^H=\sigma^H(t,H_t)$ denotes volatility.  The market price of risk and the volatility of stock could be  functions of both the stock price and the state variable, respectively; that is  $\lambda^S=\lambda^S(t,H_t,\ln{S_t})$ and $\sigma^S=\sigma^S(t,H_t,\ln{S_t})$. 
 
In this market, the number of investable risky assets is less than the number of risk drivers, hence market incompleteness. To eliminate the welfare loss resulting from the unhedgeable risk drivers, we introduce a set of financial derivatives:
 \begin{equation*}
\Omega _{O}^{(n)}=\left\{
\bar{O}_{t}=[O_{t}^{(1)},O_{t}^{(2)},...,O_{t}^{(n)}]^{T}\mid O_{t}^{(i)}\neq 0%
\text{, }i=1,...,n\text{ and rank}\left( \Sigma _{t}\right) =2, \, t\in \lbrack 0,T] \right\}.
\end{equation*}
We assume that an investor allocates in an element of $\Omega _{O}$; that is, a specific $\bar{O}_{t}=[O_{t}^{(1)},O_{t}^{(2)},...,O_{t}^{(n)}]^{T}$ $(n\geq 2)$. Note that
by arbitrage arguments, the dynamics of the extended market are as follows: 
\begin{equation}
\begin{cases}
\frac{d M_t}{M_t}=rdt\\
d\bar{O}_{t}=diag(\bar{O}_{t})\left[(r\cdot \mathbbm{1}+\Sigma _{t}\Lambda )dt+\Sigma
_{t}dB_{t}\right]\\
d H_t=\mu^Hdt+\sigma^H dB_t^H \\
<dB_t^S,dB^H_t>=\rho_{SH} dt,
\end{cases}\label{chp6_model_option}
\end{equation}
where $B_t=[B_t^S,B^H_t]^T$ and $\Sigma _{t}$ represents the $n\times2$ variance matrix of $\bar{O}_{t}$; the first column $(i,1)$ represents the sensitivity of $O_{t}^{(i)}$ to the underlying
asset $S_{t}$ (i.e. $\frac{\partial O_{t}^{(i)}}{\partial S_{t}}S_{t}\frac{1}{O_{t}^{(i)}}\sigma ^S$); and the second column $(i,2)$ represents the sensitivity of $O_{t}^{(i)}$ to the state variable $H_t$ (i.e. $\frac{\partial O_{t}^{(i)}}{\partial H_{t}}\frac{1}{O_{t}^{(i)}}\sigma ^H$). $\Lambda=[\lambda^S,\lambda^H]^T$, where $\lambda^H=\lambda^H(t,H_t,\ln{S_t})$ denotes the market price of volatility risk. Rank $2$ variance matrix $\Sigma _{t}$ guarantees the completeness of the market. For simplicity, we also assume that the derivatives in $\Omega _{O}^{(n)}$ will be rolled over, always maintaining  the same time to maturity and a non-zero value. Note that the investor is not prohibited from trading on the stock, which is included in $\Omega _{O}^{(n)}$ as a special derivative.

Let $\Omega _{\pi }^{(O)}$ denote the space of admissible strategies
satisfying the standard conditions, where the element $\pi _{t}=[\pi
_{t}^{(1)},\pi _{t}^{(2)},...,\pi _{t}^{(n)}]^{T}$ represents the proportions
of the investor's wealth in the derivatives $%
\bar{O}_{t}=[O_{t}^{(1)},O_{t}^{(2)},...,O_{t}^{(n)}]^{T}$, with the remaining $1-\mathbbm{1}%
^{T}\pi _{t}$ invested in the cash account $M_{t}$. The investor's wealth process $%
W_{t} $ satisfies
\begin{equation}
\frac{dW_{t}}{W_{t}}=(r+\pi _{t}^{T}\Sigma _{t}\Lambda )dt+\pi _{t}^T\Sigma
_{t}dB_{t}.
\end{equation}
 The investor's objective is to maximize the expected utility of their wealth at terminal $T$; hence, their problem at time $t\in[0,T]$ can be written as 
\begin{equation}
V(t,W,H,\ln{S})=\max_{\pi _{s\geq t}\in \Omega _{\pi }^{(O)}}\mathbb{E}(U(W_{T})\mid
\mathcal{F}_{t}). \label{chp6_investor_p}
\end{equation}
The associated Hamilton-Jacobi-Bellman (HJB) equation for the value function $V$ follows the principles of stochastic control and is given by
\begin{equation}
\begin{split}
 \sup_{\pi _{t}}\bigg\{ V_{t}+W_{t}V_{W}(r+\pi _{t}^{T}\Sigma _{t}\Lambda )+%
\frac{1}{2}W_{t}^{2}V_{WW}(\pi _{t}^{T}\Sigma _{t}\Phi \Phi ^{T}\Sigma
_{t}^{T}\pi _{t})+W_{t}V_{WH}\sigma^H(\pi _{t}^{T}\Sigma _{t}A)+W_{t}V_{W\ln{S}}\sigma^S(\pi _{t}^{T}\Sigma _{t}B) \bigg\} \\
+V_H \mu^H+\frac{1}{2}V_{HH}(\sigma^H)^2+V_{\ln{S}}(r+\lambda^S\sigma^S)+\frac{1}{2}V_{\ln{S}\ln{S}}(\sigma^S)^2+V_{H\ln{S}}\sigma^H\sigma^S\rho_{SH}=0, 
\end{split}  \label{chp6_hjb}
\end{equation}%
where $\Phi =\left[
\begin{array}{cc}
1 & 0 \\
\rho_{SH} & \sqrt{1-\rho_{SH} ^{2}}%
\end{array}%
\right] $, $A=[\rho_{SH},1]^T$ and $B=[1,\rho_{SH}]^T$. 

Next, we define a new artificial market, which consists of three assets: a risk-free money account $M_t$ and two pure factor assets $S_t^{(S)}$ and $S_t^{(H)}$:
\begin{equation}
\begin{cases}
\frac{d M_t}{M_t}=rdt\\
\frac{d S_t^{(S)}}{S_t^{(S)}}=(r+\lambda^S )dt+  dB_t^S\\
\frac{d S_t^{(H)}}{S_t^{(H)}}=(r+\lambda^H )dt+  dB_t^H\\
d H_t=\mu^Hdt+\sigma^H dB_t^H \\
<dB_t^S,dB^H_t>=\rho_{SH} dt.
\end{cases}\label{chp6_art_market}
\end{equation}
Compared to the original market, the market state variable is still $H_t$; nonetheless, here the investor can put their money in the hypothetical pure factor assets $S_t^{(S)}$ and $S_t^{(H)}$, which have a unit exposure on $B_t^S$ and $B_t^H$, respectively. Let $\eta_t=[\eta_t^{(1)},\eta_t^{(2)}]^T$ be the allocation on the pure factors (also known as exposures in the literature: see \cite{liu2003dynamic}); $\hat{W}_t$ denotes the investor's wealth process, and $\hat{V}(t,\hat{W},H,\ln{S})$ represents the value function in the artificial market. Similarly, the associated HJB equation is given by
\begin{equation}
\begin{split}
 \sup_{\eta _{t}}\bigg\{ \hat{V}_{t}+\hat{W}_{t}\hat{V}_{\hat{W}}(r+\eta _{t}^T\Lambda )+%
\frac{1}{2}\hat{W}_{t}^{2}\hat{V}_{\hat{W}\hat{W}}(\eta _{t}^T\Phi \Phi ^{T}\eta _{t})+\hat{W}_{t}\hat{V}_{\hat{W}H}\sigma^H(\eta _{t}^TA)+\hat{W}_{t}\hat{V}_{\hat{W}\ln{S}}\sigma^S(\eta _{t}^TB) \bigg\} \\
+\hat{V}_H \mu^H+\frac{1}{2}\hat{V}_{HH}(\sigma^H)^2+V_{\ln{S}}(r+\lambda^S\sigma^S)+\frac{1}{2}\hat{V}_{\ln{S}\ln{S}}(\sigma^S)^2+\hat{V}_{H\ln{S}}\sigma^H\sigma^S\rho_{SH}=0.
\end{split}  \label{chp6_hjb_art}
\end{equation}
If the solution of the associated HJB PDEs exists, then it is easy to verify that,  
\begin{align}
   \hat{V}(t,\hat{W},H,\ln{S})&= V(t,W,H,\ln{S}) \\
   \hat{W}_t& =W_t \\
   \Sigma_t^T\pi_t^* &= \eta_t^*. \label{chp6_eta_pi_trans}
\end{align}
Furthermore, if the number of derivatives in $O_t$ is greater than $2$ (i.e. $n\geq2$), there are infinitely many optimal strategies, all producing the same maximum value function. 

Aside from the expected utility maximization, the investor is also concerned with the size of their risky allocations. For instance, on the other hand, an institutional investor may have to keep their gross allocation exposure under a certain level due to regulatory constraints. On the other hand, a small exposure is important for capital safety regarding unmodellable risk, such as financial crisis. Hence, we consider an additional derivative selection criterion, namely risk exposure minimization, introduced in \cite{2022arXiv220103717E}: 
\begin{equation}
\underset{{\bar{O}_{t}\in \Omega _{O}^{(n)}}}{\min }\left\Vert \underset{\pi
_{s\geq t}\in \Omega _{\pi }^{(O)}}{\arg \max }\mathbb{E}(U(W_{T})\mid
\mathcal{F}_{t})\right\Vert _{1},  \label{chp6_linear_pron}
\end{equation}%
where $\left\Vert \pi _{s\geq t}\right\Vert
_{1}=\sum\limits_{i=1}^{n}\left\vert \pi _{t}^{(i)}\right\vert $ represents
the $\ell _{1}$ norm of allocations at time $t$. Note that this objective is equivalent to maximizing the cash position while shorting less. \cite{2022arXiv220103717E} demonstrated that the redundancy offers no additional help with either the investor’s expected utility or their risky asset exposure in the case of two one-factor assets. In the next proposition, we demonstrate a generalized conclusion, which applies to any diffusion model.

\begin{proposition}
Assume that an optimal solution for Problem \eqref{chp6_linear_pron} exists for $n\geq 2
$; then, \eqref{chp6_linear_pron} leads to the same minimal $\ell _{1}$ norm for any $n\geq 2$. In addition, an optimal strategy exists for Problem 
\eqref{chp6_linear_pron} such that the number of non-zero allocations is less than or equal to $2$\footnote{The result can be easily extended to higher dimension. When model contains $m\geq2$ independent risk factors (Brownian motions), an optimal strategy exists for Problem  \eqref{chp6_linear_pron} such that the number of non-zero allocations is less than or equal to $m$.}.

\begin{proof}
See Appendix \ref{chp6_appendix_twoproblem}.

\end{proof}\label{chp6_propos_twoproblem}
\end{proposition}

Proposition \ref{chp6_propos_twoproblem} demonstrates that investors do not need to consider the  portfolio composition $O_t$ with size $n>2$. Working with $n=2$ is sufficient for both Problems \eqref{chp6_investor_p} and \eqref{chp6_linear_pron}. We hence only study the simplest case given a complete market setting (i.e. $n = 2$).

\section{Polynomial affine method for CRRA utilities in financial derivatives market}
\label{chp_6_sec3}

In this section, we introduce a methodology to compute derivatives-based portfolio strategies. This method is required to find the optimal candidate composition $\bar{O}_t\in\Omega _{O}^{(2)}$ for risk exposure minimization. 

Complexity in assets' dynamic models often jeopardizes the analytic solvability of HJB PDE; this means that closed-form solutions are not always available. Motivated by this fact,  \cite{zhu2022polynomial} proposed a simulation-based method to approximate the optimal strategy for  continuous-time portfolios within EUT (i.e. the PAMC). The original PAMC method is only applicable to asset classes, such as equity, fixed income and currency, where assets' dynamics are known explicitly. However, with proper modifications, the PAMC is easily extended to financial derivatives markets. The new method, namely the PAMC-indirect, is introduced in Section \ref{chp6_indirect}. Furthermore, an alternative method is described in Appendix \ref{chp6_direct}. The performances of both methodologies are demonstrated in the case of the Heston model, and the comparison to the theoretical solution confirms the excellent accuracy and efficiency of the PAMC-indirect method.

\subsection{The PAMC-indirect}
\label{chp6_indirect}
Inspired by the quadratic affine model family (see \cite{liu2006portfolio}), the PAMC approach assumes that the value function has the following representation:
\begin{equation}
    V(t,W,H,\ln{S})=\frac{W^{1-\gamma}}{1-\gamma}f(t,H,\ln{S}),
    \label{chp6_valuefunc_ansatz}
\end{equation}
where $f(t,H,\ln{S})$ is approximated by an exponential polynomial function of order $k$; that is, $\exp\{P_k\}$. The PAMC method utilizes the Bellman equation and the fact that the value function at re-balancing time is the conditional expectation of the value function at $t+\Delta t$; that is,
\begin{equation*}
    V(t,W_t,H_t,\ln{S_t})=\max_{\pi _t}\mathbb{E}(V(t+\Delta t,W_{t+\Delta t},H_{t+\Delta t},\ln{S_{t+\Delta t}})\mid
\mathcal{F}_{t}).
\end{equation*}
The PAMC expands the value function at $t+\Delta t$ with respect to wealth $W$, state variable $H$ and log stock price $\ln{S}$, and it considers a sufficiently small re-balancing interval $\Delta t$ such that the infinitesimal $o(\Delta t)$ terms are omitted. Then, the value function $V(t,W_t,H_t,\ln{S_t})$ is rewritten as a quadratic function of the portfolio strategy, and the optimal strategy is immediately solved with the first order condition given the information at $t+\Delta t$. Proposition \ref{chp6_propos_opt} displays the estimation of optimal strategy $\eta_t^*$ in the artificial pure factor market \eqref{chp6_art_market}.

\begin{proposition}
Given the approximation of the value function at the next re-balancing time $t+\Delta t$  (i.e. $\frac{W^{1-\gamma}}{1-\gamma} \exp\{P_k\}(t+\Delta t,H,\ln{S})$), the optimal strategy at time $t$ is given by
\begin{equation}
  \eta_t^*=\frac{1}{\gamma}(\Phi \Phi ^{T})^{-1} (\Lambda + \frac{\partial P_k}{\partial H}\sigma^HA+\frac{\partial P_k}{\partial \ln{S}}\sigma^SB ).
  \label{chp6_opt_pi_indirect}
\end{equation}

\begin{proof}
See Appendix \ref{chp6_appendix_opt}.

\end{proof}\label{chp6_propos_opt}
\end{proposition} 
The PAMC-indirect inherits the recursive approximation structure of the PAMC. After the generation of paths of asset price and state variables, the optimal pure factor strategies at last re-balancing time $T-\Delta t$ can be directly computed with \eqref{chp6_opt_pi_indirect} because $P_k(T,H,\ln{S})=0$; the path-wise expected utilities are obtained through simulation. Furthermore, the expected utilities are regressed over stock price $S_{T-\Delta t}$ and state variable $H_{T-\Delta t}$, and the regression function approximates the $V(T-\Delta t,W,H,\ln{S})$. Then, the method moves backward, and similar procedures are conducted at each re-balancing time until the optimal initial strategy of the pure factor portfolio (i.e. $\eta_0^*$) is obtained.

Finally, the PAMC-indirect calculates the portfolio variance matrix $\Sigma_t$, which depends on the option price $O_t$, Delta $\frac{\partial O_t}{\partial S_t}$ and the sensitivity to the state variable $\frac{\partial O_t}{\partial H_t}$. The optimal derivatives strategy $\pi_0^*$ is solved with \eqref{chp6_eta_pi_trans}. Only in some special cases (e.g. the Black-Scholes model) are option prices solved analytically. A variety of approximation methods for option price and Greeks are available in the existing literature. The choice of such methods should be determined by the option style and underlying assets model. For example, an accurate Fourier transform (FT) approximation is an ideal choice when the semi-closed-form solution of an option is available (e.g. the Heston model, the Ornstein–Uhlenbeck 4/2 model), while a simple Monte Carlo simulation is universal for options with a deterministic exercise date; and a least-squares Monte Carlo method is applicable when considering American style options. 

We clarify the notation in Table \ref{chp6_note} and detail the PAMC-indirect in Algorithm \ref{chp6_PAMH_indirect}.

\begin{table}[h]
\centering
\begin{tabular}{|ll|}
\hline
Notation & Meaning  \\
\hline
 $B_t^{m,S}$,$B_t^{m,H}$        & Brownian motion at time $t$ in $m^{th}$ simulated path      \\
 $S_t^{m}$       & Stock price at time $t$ in $m^{th}$ simulated path          \\
 $S_t^{m,S}$       &  Pure factor asset $S_t^{S}$  at time $t$ in $m^{th}$ simulated path          \\
 $S_t^{m,H}$       & Pure factor asset $S_t^{H}$ at time $t$ in $m^{th}$ simulated path          \\

$H_t^m$       & State variable Stock price at time $t$ in $m^{th}$ simulated path         \\
$O_t^m$       & Derivatives price  at time $t$ in $m^{th}$ simulated path     \\
 $n_r$        & Number of simulated paths      \\
  $\Sigma_t^m$        & Variance matrix of portfolio composition at time $t$ in $m^{th}$ simulated path    \\
 $N$        & Number of simulations to compute expected utility for a given set $(W_0,S_t^{m},H_t^m)$ \\
  $\hat{W}_{t+{\Delta t}}^{m,n}(\pi^m)$ &  The simulated wealth level at $t+{\Delta t}$ given the wealth, the allocation and other state variables\\& at $t$ are $W_0$, $\pi^m$, $S_{t}^m$, and  $H_{t}^m$   \\
$\hat{S}_{t+{\Delta t}}^{m,n}$   &  A simulated stock price at $t+{\Delta t}$ given $S_t^{m}$   \\
$\hat{H}_{t+{\Delta t}}^{m,n}$   &  A simulated state variable at $t+{\Delta t}$ given $H_t^m$  \\
$\hat{O}_{t+{\Delta t}}^{m,n}$   &  A simulated option price at $t+{\Delta t}$   \\
 $V(t,W,\ln{S},H)$           &   Value function at time $t$ given wealth $W$, stock price $S$ and state variable $H$       \\

 $\hat{v}^{m}$ &Estimation of $P_k(t,\ln{S_t^m},H_t^m)= \log(f(t,\ln{S_t^m},H_t^m))$ in \eqref{chp6_valuefunc_ansatz}.  Regressand in regression; \\ & superscript ${m}$ indicates the corresponding regressor  $(\ln{S_t^m},H_t^m)$\\
$L_{t}(H,\ln{S})$  & The regression function to be used to approximate  $P_k(t,\ln{S},H)$  \\
$\eta_{t}^{m}$   & Optimal strategy at time $t$ in $m^{th}$ simulated path   \\ \hline
\end{tabular}\caption{Notation and definitions}\label{chp6_note}
\end{table}

\begin{algorithm}[h]
\caption{PAMC-indirect} \label{chp6_PAMH_indirect}
      \KwIn{$S_0$,$W_0$,$H_0$\\}
      \KwOut{Optimal trading strategy $\pi^*_0$}
      Initialization\;
      Generating $n_r$ paths of $B_t^{m,S}$, $B_t^{m,H}$,$S_t^{m}$, $H_t^m$,$S_t^{m,S}$,$S_t^{m,H}$ $ \quad for \quad m= 1 ...n_r$\;
     
      \While{ $t=T-\Delta t$ }{
      \For{$m= 1 ...n_r$}{
        Directly compute optimal allocation $\eta^m_{T-\Delta t}$ with Equation \eqref{chp6_opt_pi_indirect} where $P_k=1$ at time $T$\;
        \For{ $n= 1 ...N$}{ 
        Generate $\hat{S}_T^{m,n,S}$ and $\hat{S}_T^{m,n,H}$ given $S_{T-\Delta}^{m,S}$ and $S_{T-\Delta}^{m,H}$ \;}
        Compute wealth $\hat{W}_T^{m,n}(\eta^m_{T-\Delta t})$ at the terminal time given the wealth at $W_{T-\Delta t}=W_0$, the transformed value function is estimated by
        $\hat{v}^{m}=\ln{[(1-\gamma)\frac{1}{N}\sum\limits_{n=1}^{N}U(\hat{W}_T^{m,n}(\pi^m_{T-\Delta t}))]}-(1-\gamma)\ln{W_0}$ \;}
        Regress $\hat{v}^{m}$ over the polynomial of $H_{T-\Delta t}^m$ and $\ln{S}_{T-\Delta t}^m$, and obtain the function $L_{T-\Delta t}(H,\ln{S})$\; }

       \For{$t=T-2\Delta t$ to $\Delta t$}
        {
         \For{$m= 1 ...n_r$}{
        Directly compute optimal allocation $\eta^m_{t}$ with Equation \eqref{chp6_opt_pi_indirect} where $P_k=L_{t+\Delta t}(H,\ln{S})$\;
         \For{ $n= 1 ...N$}{ 
        Generate $\hat{S}_{t+\Delta t}^{m,n}$, $\hat{H}_{t+\Delta t}^{m,n}$, $\hat{S}_{t+\Delta t}^{m,n,S}$ and $\hat{S}_{t+\Delta t}^{m,n,H}$ given $S_{t}^{m}$, $H_{t}^{m}$, $S_{t}^{m,S}$ and $S_{t}^{m,H}$  \;}
        Compute wealth $\hat{W}_{T+\Delta t}^{m,n}(\eta^m_{t})$ at the terminal given the wealth at $W_t=W_0$, the transformed value function is estimated by
         $\hat{v}^{m}=\ln{[\frac{1}{N}\sum\limits_{n=1}^{N}(W^{m,n}_{t+\Delta t}(\pi^m_t))^{1-\gamma}exp(L_{t+\Delta t}(\hat{H}_{t+\Delta t}^{m,n},\ln{\hat{S}}_{t+\Delta t}^{m,n}))]} -(1-\gamma)\ln{W_0}$  \;}
        Regress $\hat{v}^{m}$ over the polynomial of $H_{t}^m$ and $\ln{S}_{t}^m$, and obtain the function $L_{t}(H,\ln{S})$\; }
        
        \While{ $t=0$ }{ 
        $\eta_0^*$ is obtained with Equation \eqref{chp6_opt_pi_indirect} and where the $P_k=L_{\Delta t}(H,\ln{S})$\;
        Apply approximation methods and obtain the price of  $O_0(H_0,\ln{S_0})$ as well as its sensitivity $\frac{\partial O_0}{\partial S_0}(H_0,\ln{S_0})$ and $\frac{\partial O_0}{\partial H_0}(H_0,\ln{S_0})$ \;
        Compute the variance matrix $\Sigma_0$, and the optimal allocation $\pi_0^* =(\Sigma_t^T)^{-1} \eta_0^*$\;
       
        }

         return $\pi^*_{0}$
\end{algorithm}

\section{Derivatives selection}
\label{chp_6_sec4}

In this section, we study derivative selection for market completion-that is, \eqref{chp6_linear_pron}-for $n=2$ within subsets of the derivative set $\Omega_O^{(2)}$. The derivative selection problem is rewritten as 
\begin{equation}
\underset{{\bar{O}_{t}\in \Omega _{O}^{(2,C)}}}{\min }\left\Vert \underset{\pi
_{s\geq t}\in \Omega _{\pi }^{(O)}}{\arg \max }\mathbb{E}(U(W_{T})\mid
\mathcal{F}_{t})\right\Vert _{1},  \label{chp6_linear_pron1}
\end{equation}
where $\Omega _{O}^{(2,C)}$ is a derivative set defined by
\begin{equation*}
  \Omega _{O}^{(2,C)}=\left\{\bar{O}_t=[S_t, O_t^{(C)}]^T \vert O_t^{(C)}\in C, t\in[0,T]\right\}.
\end{equation*}
The portfolio composition $\bar{O}_t\in\Omega _{O}^{(2,C)}$ consists of a stock $S_t$ and a derivative security $O_t^{(C)}$; superscript $C$ represents the candidate set of derivative type; and $T_{op}$ denotes the time to maturity of $O_t^{(C)}$. This setting coincides with a popular practical strategic investment implementation (i.e. the elimination unhedgeable risk factors of a pure-stock portfolio with financial derivatives). We use the Heston SV model given in \eqref{chp6_heston} as the proxy of the market dynamics.

\begin{equation}
\begin{cases}
\frac{d M_t}{M_t}=rdt\\
\frac{d S_t}{S_t}=(r+\lambda X_t)dt+\sqrt{X_t} dB_t^S\\
dX_t=\kappa^X(\theta^X-X_t)dt+\sigma^X\sqrt{X_t}dB_t^{X}\\
dO^{(C)}_t=(r+\lambda\frac{\partial O^{(C)}_t}{\partial  S_t}S_tX_t+\lambda^X\frac{\partial O^{(C)}_t}{\partial  X_t}\sigma^XX_t)dt+\frac{\partial O^{(C)}_t}{\partial  S_t}S_t\sqrt{X_t}dB_t^{S}+\frac{\partial O^{(C)}_t}{\partial  X_t}\sigma^X\sqrt{X_t}dB_t^{X}\\	<B^S,B^{X}>_t=\rho_{S{X}}
\end{cases}\label{chp6_heston}
\end{equation}
The Heston model is a specific case of the generalized diffusion model \eqref{chp6_gene_diff_model} with $\lambda^S=\lambda\sqrt{X_t}$, $\lambda^H=\lambda^X\sqrt{X_t}$, $\sigma^S=\sqrt{X_t}$, $\mu^X=\kappa^X(\theta^X-X_t)$ and $\sigma^H=\sigma^X\sqrt{X_t}$. We employed a representative market-calibrated set of parameters (see Table \ref{chp6_para_heston}), given in \cite{liu2003dynamic}, to investigate the best product to account for volatility risk. The optimal allocation for the model \eqref{chp6_heston} can be written explicitly with Equations \eqref{chp6_opt_pi_indirect} and \eqref{chp6_eta_pi_trans} as follows
\begin{eqnarray}
\pi^S_t & = & \frac{1}{\gamma(1-\rho_{SX}^2)}(\lambda-\rho_{SX}\lambda^X)-\pi^O_t\frac{S_t}{O^{(C)}_t}\frac{\partial O^{(C)}_t}{\partial S_t}\notag \\
\pi^O_t & = & \left(\frac{O^{(C)}_t}{\gamma\sigma^X (1-\rho_{SX}^2)}(\lambda^X-\rho_{SX}\lambda)+\frac{O^{(C)}_t}{\gamma}\frac{\partial P_k}{\partial X_t} \right)\frac{1}{\frac{\partial O^{(C)}_t}{\partial X_t}}.
\label{chp6_pi_explicit}
\end{eqnarray}
The representation indicates that the optimal allocation on option $\pi^O_t$ solely depends on the choice of option $O_t^{(C)}$ (i.e. $\pi^O_t$ is a function of the option's sensitivity to the instantaneous variance and option price), while the optimal allocation on the stock $\pi^S_t$ is determined by the ratio of the option's sensitivity to the instantaneous variance and the sensitivity to the stock.

\begin{table}[H]\centering\caption{Parameter value for the Heston model.}\label{chp6_para_heston}
\begin{tabular}{|llll|}\hline
Parameter &Value& Parameter &Value \\ \hline
$T$& 1 year &$\rho_{SX}$&-0.4  \\
$\theta^X$  & 0.0169  &   $\sigma^X$  &  0.25\\
$\kappa^X$ & 5.0  &$\lambda$ & 4.0 \\
$\lambda^X$ & -7.1 & $T_{op}$ & 0.1 year\\
$\Delta t$ & $\frac{1}{60}$ & $period$ & 60 \\
$r$ & 0.05 & $X_0$ & $\theta^X$ \\
$S_0$ & 1.0 & $M_0$ & 1.0  \\
$W_0$ & 1 & $\gamma$ & 4\\
 $N$ & 2000 & $n_r$  & 100\\
 \hline
\end{tabular}
\end{table}
\subsection{Derivatives selection within  options on stock}
\label{chp6_4.1}
We start the selection among four popular equity options. Specifically, the candidate set is given by
\begin{equation*}
    C=\left\{ \textit{Call option, Put option, Straddle, Strangle}\right\}.
\end{equation*}
 
For simplicity, we only consider European-style derivatives. Call (i.e. payoff $(S-K)^+$) and put (i.e. payoff $(K-S)^+$) options are the most common products traded in the market. Additionally, a straddle (i.e. payoff $(S-K)^+ + (K-S)^+$) is a commonly used product when investors expect the underlying asset to deviate from the spot price; hence, the long position of a straddle is approximately a long position on volatility. Compared with a straddle synthesized by purchasing a call and a put with the same strike price and maturity, a strangle (i.e. payoff $(S-K_1)^+ + (K_2-S)^+$) has a more flexible structure, as it takes long positions on out-of-the-money (OTM) put and call, which is a cheaper way to acquire exposure to volatility \footnote{Elements in variance matrix $\Sigma_t$, which are functions of option prices and Greeks, can be obtained with numerical integration method (see \cite{rouah2013heston} chapter 11). Specifically, we utilized the formula given in \cite{heston1993closed} and compute numerical integration with the Newton-Cotes formulas.   }.

\begin{figure}[H]
	\centering
	\vspace{0.35cm}
	\subfigtopskip=2pt
	\subfigbottomskip=2pt
	\subfigcapskip=-5pt
	\subfigure[Maturity=0.1]{
		\includegraphics[width=0.45\linewidth]{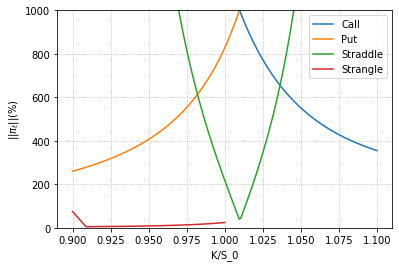}}
	\subfigure[Maturity=0.5]{
		\includegraphics[width=0.45\linewidth]{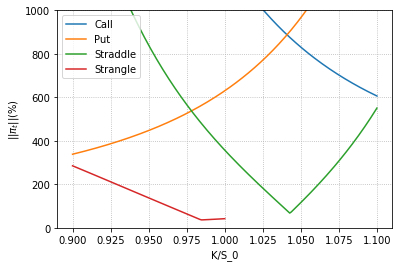}}

	\caption{The $\Vert\pi_t \Vert_1 $ versus moneyness: The Y-axis is the risk exposure of a portfolio containing different derivatives. The X-axis indicates the moneyness $K/S_0$ of calls, puts and straddles.\\ 
	The strangle is synthesized with an OTM put and an OTM call. Given moneyness of the OTM put indicated by the X-axis, the strike price of the OTM call is the one achieving minimum $\Vert\pi_t \Vert _1$ within the range $[S_0,110\%S_0]$. }
    \label{chp6_pi_heston_stock}
\end{figure}

Figure \ref{chp6_pi_heston_stock} displays the risk exposure $\Vert\pi_t \Vert _1$ of portfolios  as a function of derivative moneyness $K/S_0$, where $K$ is the strike price of the options. Figure \ref{chp6_pi_heston_stock} (a) exhibits risk exposure given options with maturity $T_{op}=0.1$, and Figure  \ref{chp6_pi_heston_stock} (b) displays results when the option maturity is $T_{op}=0.5$. In both cases, investors reduce their risk exposure with OTM put and call options. Puts and calls could lead to illiquid choices, whereas a straddle achieves minimum $\Vert\pi_t \Vert _1$ when it is near at-the-money (ATM). The optimal moneyness of a straddle option shifts to the right as maturity $T_{op}$ increases. The risk exposure with a strangle decreases as its component put option moves deeper OTM. Furthermore, even the strangle consisting of a near-ATM put and call outperforms other options. We consequently conclude that the strangle minimizes the risk exposure.

\begin{figure}[H]
	\centering
	\vspace{0.35cm}
	\subfigtopskip=2pt
	\subfigbottomskip=2pt
	\subfigcapskip=-5pt
	\subfigure[Maturity=0.1 ]{
		\includegraphics[width=0.45\linewidth]{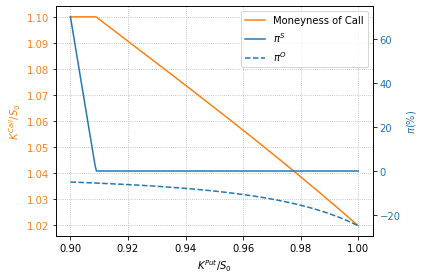}}
    \subfigure[Maturity=0.5]{
		\includegraphics[width=0.45\linewidth]{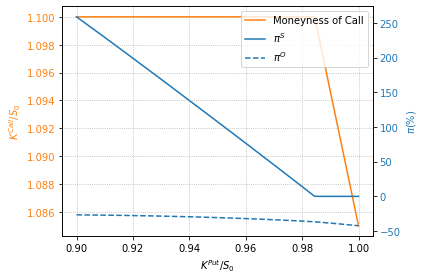}}

	\caption{Impact of the OTM put's moneyness on the strangle. Left vertical axis indicates the optimal moneyness of the OTM call within $[100\%S_0,110\%S_0]$. Right vertical axis indicates the allocation on the stock and the strangle.}
    \label{chp6_pi_heston_stran_turn}
\end{figure}

The turning point on the left tail of the strangle's risk exposure in Figure \ref{chp6_pi_heston_stock} is further studied in Figure \ref{chp6_pi_heston_stran_turn}, where we illustrate how the optimal moneyness of an OTM call, an allocation on stock $\pi_t^S$ and an allocation on strangle $\pi_t^O$ vary with the moneyness of an OTM put. Note the practical range selected for the moneyness of an OTM call; that is, $K^{Call}/S_0\in[S_0,110\%S_0]$. It is shown that, if the strike price of the put option starting at the spot price moves in the direction of OTM, the corresponding optimal moneyness of the call option also becomes deeply OTM. The OTM call reaches the boundary earlier than the put, which leads to the turning point. Before the turning point, allocation on the stock $\pi^S_t$ continues to be small, and $\pi^O_t$ gradually approaches $0$; hence, the total risk exposure $\Vert\pi_t\Vert_1$ assumes a decreasing trend. However, $\pi^S_t$ increases rapidly after the turning point, and $\Vert\pi_t\Vert_1$ consequently rises as $\pi^O_t$ continues to drop. Moreover, Figures \ref{chp6_pi_heston_stran_turn} (a) and (b) compare strangles with maturity $T_{op}=0.1$ and $T_{op}=0.5$, respectively. The turning point for a longer maturity strangle is more easily reached, which makes it less preferable.

\begin{figure}[H]
	\centering
	\vspace{0.35cm}
	\subfigtopskip=2pt
	\subfigbottomskip=2pt
	\subfigcapskip=-5pt
	\subfigure[Vega $\frac{\partial O^{(C)}_t}{\partial X_t}$ ]{
		\includegraphics[width=0.45\linewidth]{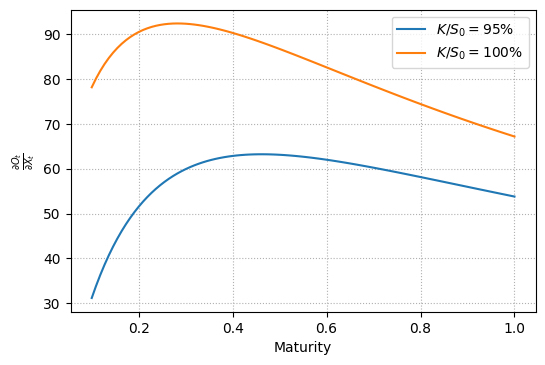}}
    \subfigure[Vega to Price $\frac{\partial O^{(C)}_t}{\partial X_t}/O^{(C)}_t$ ]{
		\includegraphics[width=0.45\linewidth]{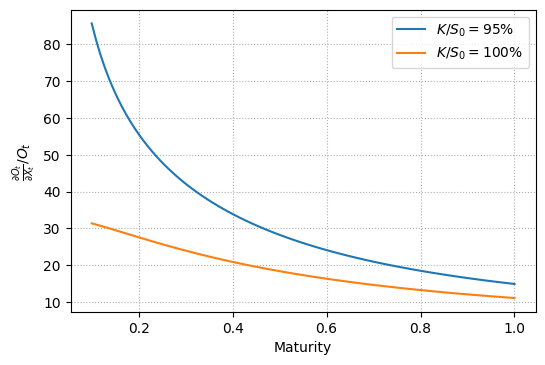}}

	\caption{Sensitivity of strangle option price $O^{(C)}_t$ to instantaneous variance $X_t$ versus time to maturity $T_{op}$. The legend indicates the moneyness of component put, and the call option is the one achieving minimum risk exposure. Note that there is no boundary for the strike price of the component call. }
    \label{chp6_pi_heston_vega}
\end{figure}

Equation \eqref{chp6_pi_explicit} demonstrates that the allocation on the option is determined by the ratio of the Vega to the option price. Therefore, in Figure \ref{chp6_pi_heston_vega}, we investigate the relationship between the Vega of the strangle and the time to maturity to provide further insight for the comparison of maturity in Figures \ref{chp6_pi_heston_stock} and \ref{chp6_pi_heston_stran_turn}. Figure \ref{chp6_pi_heston_vega} (a) illustrates the Vega versus the maturity of an ATM strangle (the moneyness of component put option $K/S_0=100\%$) and an OTM strangle (the moneyness of component put option $K/S_0=95\%$). For an especially short-term maturity strangle, the terminal payoff do not have sufficient time to react to the change in volatility state, therefore, the Vega is small. For the long-term maturity strangle, a change in the instantaneous variance also has a small impact on the option price because of its mean-reverting nature. Hence, the Vegas of both strangles are concave in time to maturity, which peaks at around $0.3$ years. The impact from time to maturity on the ratio of Vega to price is illustrated in Figure \ref{chp6_pi_heston_vega} (b), where $\frac{\partial O^{(C)}_t}{\partial X_t}/O^{(C)}_t$ is always positive and monotonically decreases with maturity, which leads to an increasing $\vert\pi_t^O\vert$. In Figure \ref{chp6_pi_heston_stran_turn}, $\pi_t^S$ is close to $0$ before the boundary, and  $\vert\pi_t^O\vert$ increases with maturity; hence, we conclude that a short-term maturity strangle is preferable.

\subsection{Derivatives selection within VIX products}
 Next, we study an investor who has access to the VIX of the stock at hand, such as the VIX for the S\&P 500. In this case, the investor has direct access to the volatility risk by investing in products based on the VIX. The VIX has drawn investors' attention since its origin in 1993; not only is it a real-time indicator of the market sentiment, but also products such as VIX futures and VIX options are popular for hedging volatility risk. In this section, we explore products on the VIX. We consider a candidate set
\begin{equation*}
    C=\left\{ \textit{VIX call, VIX put, VIX straddle, Strangle}\right\}.
\end{equation*}
Note that a strangle is the best option for minimizing risk exposure considered in Section \ref{chp6_4.1}.  VIX calls and VIX puts are call and put options, respectively, based on the value of the VIX. A VIX straddle is an instrument synthesized by the long position of a VIX call and a VIX put with the same strike price.

Given the definition of VIX as specified in the CBOE white paper \cite{cboe2003cboe}, \cite{lin2007pricing} solved the VIX$^2$ in closed-form as a function of instantaneous variance $X_t$. Under the Heston model, we have
\begin{equation}
\begin{split}
        \textit{VIX}^2_t&=\frac{1}{\tau}\left(a_{\tau}X_t+b_{\tau}\right)\\
a_{\tau}=\frac{1-\exp{-\kappa_v^*\tau}}{\kappa_v^*}, \quad b_{\tau}=\theta_v^*(\tau-a_{\tau})\quad &\kappa_v^*=\kappa_v+\lambda^X\sigma_v,  \quad \theta_v^*=\frac{\kappa_v\theta_v}{\kappa_v^*}, \quad \tau=\frac{30}{365},
\end{split}
\end{equation}
where VIX$^2_t$ is linear with the instantaneous variance $X_t$. Computing a VIX option's price and Greeks is easy via Monte Carlo simulation; this method enable us to find elements in the variance matrix $\Sigma_t$. 

Unlike options on the stock, by investing in VIX products, the investor acquires exposure only on the volatility risk; hence, the variance matrix $\Sigma_t$ is diagonal. Moreover, the equity-neutral position of VIX products leads to a specific case of \eqref{chp6_pi_explicit}:
\begin{eqnarray}
\pi^S_t & = & \frac{1}{\gamma(1-\rho_{SX}^2)}(\lambda-\rho_{SX}\lambda^X) \notag \\
\pi^O_t & = & \left(\frac{O^{(C)}_t}{\gamma\sigma^X (1-\rho_{SX}^2)}(\lambda^X-\rho_{SX}\lambda)+\frac{O^{(C)}_t}{\gamma}\frac{\partial P_k}{\partial X_t} \right)\frac{1}{\frac{\partial O^{(C)}_t}{\partial X_t}}.
\label{chp6_vix_lower}
\end{eqnarray}
In this case, the allocation on the stock is invariant to the choice of VIX products, which thus becomes a natural lower bound for risk exposure (i.e. $\Vert\pi_t \Vert_1\geq \vert \pi_t^S\vert $ ).

\begin{figure}[H]
	\centering
	\vspace{0.35cm}
	\subfigtopskip=2pt
	\subfigbottomskip=2pt
	\subfigcapskip=-5pt
	\subfigure[{The $\Vert\pi_t \Vert_1$ versus moneyness: the time to maturity of both VIX options and the strangle is $0.1$ year. The X-axis indicates moneyness of VIX options and the OTM put in the strangle; the strike price of the OTM call is the one achieving minimum $\Vert\pi_t \Vert_1$ within the range  $\lbrack S_0,105\%S_0\rbrack$. }]{
		\includegraphics[width=0.45\linewidth]{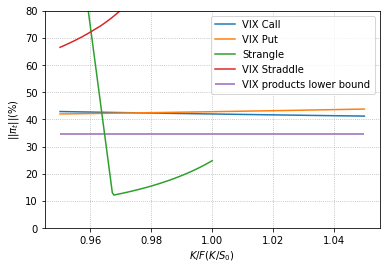}}
    \subfigure[{The $\Vert\pi_t \Vert_1 $ versus maturity: the strike price of VIX calls is $105\%S_0$. The strike price of VIX puts is $95\%S_0$. The green line shows the smallest $\Vert\pi_t \Vert_1 $ is achieved by the strangle given the OTM put strike price $K^{Put}\in[95\%S_0,100\%S_0]$ and the OTM call strike price $K^{Call}\in[100\%S_0,105\%S_0]$. } ]{
		\includegraphics[width=0.45\linewidth]{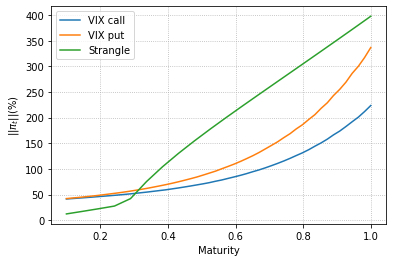}}

	\caption{The $\Vert\pi_t \Vert_1 $ for VIX products}
    \label{chp6_pi_heston_new}
\end{figure}

The risk exposure when investors hedge the volatility risk with VIX calls and puts is displayed in Figure \ref{chp6_pi_heston_new} (a). On the one hand, calls and puts on the VIX have similar properties as those on the stock: OTM options tend to achieve smaller risk exposure. On the other hand, a VIX straddle is less ineffective in hedging the volatility risk because it is relatively insensitive to the volatility, and a larger risk exposure $\Vert\pi_t \Vert_1$ is needed for investors compared to the cases of VIX calls and puts. The risk exposure with the equity strangle is displayed for comparison purpose; here, the turning point resulting from the boundary of moneyness on the OTM call is still evident. Moreover, the strangle achieves a much smaller risk exposure than the VIX products. We therefore conclude that equity strangle is superior when the time to maturity $T_{op}$ for candidate products is small ($T_{op}=0.1$). 

Figure \ref{chp6_pi_heston_new} (b) illustrates how the option maturity $T_{op}$ affects the risk exposure $\Vert\pi_t \Vert_1$. It indicates that an OTM VIX call and an OTM VIX put are preferable in (a), and a similar conclusion is verified numerically for any $T_{op}\in(0,1]$. Therefore, risk exposure for the best VIX call ($K=105\%S_0$) and VIX put ($K=95\%S_0$) are plotted in Figure \ref{chp6_pi_heston_new} (b). In addition, the  minimum risk exposure within a pre-specified region of moneyness is also displayed. As the volatility time series exhibits a mean-reverting property, the VIX options with long-term maturity are insensitive to the instantaneous variance; hence, it has little effect in hedging the volatility risk. The figure also suggests that a large allocation on the long-term maturity VIX option is needed, such that the risk exposure increases rapidly with maturity. A strangle achieves smaller risk exposure when short-term maturity products are available in the market, aligning with the result in Figure \ref{chp6_pi_heston_new}  (a).  

According to Figures \ref{chp6_pi_heston_stock} and \ref{chp6_pi_heston_stran_turn}, the boundary of the OTM call is reached faster as $T_{op}$ increases, and the boundary significantly restricts risk exposure, thus  reducing the effect of the strangle. This leads to a steep slope of risk exposure for the strangle in Figure \ref{chp6_pi_heston_new} (b). In summary, the investor should make a choice between VIX products and an equity strangle, depending on the situation. If the investor has access to short-term maturity options, then the strangle is preferable. However, when only long-term maturity products are available, the investor should choose call options on the VIX for market completion.

\section{Conclusion}
\label{chp_6_sec5}
This paper explored optimal derivatives-based portfolios to complete a market characterized by volatility risk as a state variable. An accurate and high-speed approximation for optimal allocations is proposed, for the unsolvable problem of optimal derivative exposure. In addition to the traditional portfolio decision objective (i.e. EUT maximization),  we work with an additional criterion, namely risk exposure minimization, for derivative selection. This aids in the selection of a meaningful product out of many that maximize the utility. We found that strangle options are the best equity option product for managing volatility risk. Moreover, we demonstrated that options based on the VIX are superior to equity strangles in some realistic situations. 

There are many interesting potential extensions to this line of research. For instance, we could incorporate multi-factor models considering the stochastic interest rates, stochastic correlations, jumps and stochastic market prices of risk, to mention a few. These are more realistic settings, solvable within our numerical method, hence providing investors with valuable insight into optimal high-dimensional portfolios and multi-asset derivatives for sensible practical investment.

\bibliographystyle{QF}

\bibliography{bib}

\begin{thebibliography}{25}
\providecommand{\natexlab}[1]{#1}

\bibitem[{Arrow(1964)}]{arrow1964role}
Arrow, K.~J.
\newblock The role of securities in the optimal allocation of risk-bearing.
\newblock \emph{The review of economic studies}, \textbf{31}, 2, (1964),
  91--96.

\bibitem[{Arrow and Debreu(1954)}]{arrow1954existence}
Arrow, K.~J. and Debreu, G.
\newblock Existence of an equilibrium for a competitive economy.
\newblock \emph{Econometrica: Journal of the Econometric Society}, 265--290.

\bibitem[{Bertsimas and Tsitsiklis(1997)}]{bertsimas1997introduction}
Bertsimas, D. and Tsitsiklis, J.~N.
\newblock \emph{Introduction to linear optimization}, volume~6 (Athena
  Scientific Belmont, MA), 1997.

\bibitem[{Brandt et~al.(2005)Brandt, Goyal, Santa-Clara, and
  Stroud}]{brandt2005simulation}
Brandt, M.~W., Goyal, A., Santa-Clara, P., and Stroud, J.~R.
\newblock A simulation approach to dynamic portfolio choice with an application
  to learning about return predictability.
\newblock \emph{The Review of Financial Studies}, \textbf{18}, 3, (2005),
  831--873.

\bibitem[{CBOE(2003)}]{cboe2003cboe}
CBOE, C.
\newblock Cboe vix white paper.
\newblock \emph{https://cdn.cboe.com/resources/vix/vixwhite.pdf}.

\bibitem[{Chen et~al.(2011)Chen, Chung, and Ho}]{chen2011diversification}
Chen, H.-C., Chung, S.-L., and Ho, K.-Y.
\newblock The diversification effects of volatility-related assets.
\newblock \emph{Journal of Banking \& Finance}, \textbf{35}, 5, (2011),
  1179--1189.

\bibitem[{Cheng and Escobar-Anel(2021)}]{cheng2021optimal}
Cheng, Y. and Escobar-Anel, M.
\newblock Optimal investment strategy in the family of 4/2 stochastic
  volatility models.
\newblock \emph{Quantitative Finance}, 1--29.

\bibitem[{Cong and Oosterlee(2017)}]{cong2017accurate}
Cong, F. and Oosterlee, C.~W.
\newblock Accurate and robust numerical methods for the dynamic portfolio
  management problem.
\newblock \emph{Computational Economics}, \textbf{49}, 3, (2017), 433--458.

\bibitem[{Doran(2020)}]{doran2020volatility}
Doran, J.~S.
\newblock Volatility as an asset class: Holding vix in a portfolio.
\newblock \emph{Journal of Futures Markets}, \textbf{40}, 6, (2020), 841--859.

\bibitem[{Escobar et~al.(2017)Escobar, Ferrando, and
  Rubtsov}]{escobar2017optimal}
Escobar, M., Ferrando, S., and Rubtsov, A.
\newblock Optimal investment under multi-factor stochastic volatility.
\newblock \emph{Quantitative Finance}, \textbf{17}, 2, (2017), 241--260.

\bibitem[{{Escobar-Anel} et~al.(2022){Escobar-Anel}, {Davison}, and
  {Zhu}}]{2022arXiv220103717E}
{Escobar-Anel}, M., {Davison}, M., and {Zhu}, Y.
\newblock {Derivatives-based portfolio decisions. An expected utility insight}.
\newblock \emph{arXiv e-prints}, arXiv:2201.03717.

\bibitem[{Grasselli(2017)}]{grasselli20174}
Grasselli, M.
\newblock The 4/2 stochastic volatility model: a unified approach for the
  heston and the 3/2 model.
\newblock \emph{Mathematical Finance}, \textbf{27}, 4, (2017), 1013--1034.

\bibitem[{Heston(1993)}]{heston1993closed}
Heston, S.~L.
\newblock A closed-form solution for options with stochastic volatility with
  applications to bond and currency options.
\newblock \emph{The review of financial studies}, \textbf{6}, 2, (1993),
  327--343.

\bibitem[{Heston(1997)}]{heston1997simple}
Heston, S.~L.
\newblock A simple new formula for options with stochastic volatility.

\bibitem[{Jain and Oosterlee(2015)}]{jain2015stochastic}
Jain, S. and Oosterlee, C.~W.
\newblock The stochastic grid bundling method: Efficient pricing of bermudan
  options and their greeks.
\newblock \emph{Applied Mathematics and Computation}, \textbf{269}, (2015),
  412--431.

\bibitem[{Li et~al.(2018)Li, Shen, and Zeng}]{li2018dynamic}
Li, D., Shen, Y., and Zeng, Y.
\newblock Dynamic derivative-based investment strategy for mean--variance
  asset--liability management with stochastic volatility.
\newblock \emph{Insurance: Mathematics and Economics}, \textbf{78}, (2018),
  72--86.

\bibitem[{Lin(2007)}]{lin2007pricing}
Lin, Y.-N.
\newblock Pricing vix futures: Evidence from integrated physical and
  risk-neutral probability measures.
\newblock \emph{Journal of Futures Markets: Futures, Options, and Other
  Derivative Products}, \textbf{27}, 12, (2007), 1175--1217.

\bibitem[{Liu(2006)}]{liu2006portfolio}
Liu, J.
\newblock Portfolio selection in stochastic environments.
\newblock \emph{The Review of Financial Studies}, \textbf{20}, 1, (2006),
  1--39.

\bibitem[{Liu and Pan(2003)}]{liu2003dynamic}
Liu, J. and Pan, J.
\newblock Dynamic derivative strategies.
\newblock \emph{Journal of Financial Economics}, \textbf{69}, 3, (2003),
  401--430.

\bibitem[{Longstaff and Schwartz(2001)}]{longstaff2001valuing}
Longstaff, F.~A. and Schwartz, E.~S.
\newblock Valuing american options by simulation: a simple least-squares
  approach.
\newblock \emph{The review of financial studies}, \textbf{14}, 1, (2001),
  113--147.

\bibitem[{Merton(1969)}]{merton1969lifetime}
Merton, R.~C.
\newblock Lifetime portfolio selection under uncertainty: The continuous-time
  case.
\newblock \emph{The review of Economics and Statistics}, 247--257.

\bibitem[{Rardin and Rardin(1998)}]{rardin1998optimization}
Rardin, R.~L. and Rardin, R.~L.
\newblock \emph{Optimization in operations research}, volume 166 (Prentice Hall
  Upper Saddle River, NJ), 1998.

\bibitem[{Rouah(2013)}]{rouah2013heston}
Rouah, F.~D.
\newblock \emph{The Heston model and its extensions in Matlab and C} (John
  Wiley \& Sons), 2013.

\bibitem[{Warren(2012)}]{warren2012can}
Warren, G.~J.
\newblock Can investing in volatility help meet your portfolio objectives?
\newblock \emph{The Journal of Portfolio Management}, \textbf{38}, 2, (2012),
  82--98.

\bibitem[{Zhu and Escobar-Anel(2022)}]{zhu2022polynomial}
Zhu, Y. and Escobar-Anel, M.
\newblock Polynomial affine approach to hara utility maximization with
  applications to ornsteinuhlenbeck 4/2 models.
\newblock \emph{Applied Mathematics and Computation}, \textbf{418}, (2022),
  126836.

\end{thebibliography}

\newpage

\appendix

\section{Proofs}
\label{chp6_appendix_proof}
\subsection{Proof of Proposition \ref{chp6_propos_twoproblem}}   
\label{chp6_appendix_twoproblem}
Let $O_{t,n}=[O^{(1)}_t, O^{(2)}_t,..., O^{(n)}_t]^T$ with variance matrix $\Sigma_t$ of rank 2 be an optimal subset of options for problem \eqref{chp6_linear_pron}.  $\pi^*_{t,n}$ is a strategy maximizing the expected utility if and only if $\Sigma^T_t\pi^*_{t,n}=\eta^*_t $. Therefore, $O_{t,n}$ and $\pi^*_{t,n}$ is an optimal pair for \eqref{chp6_linear_pron} when $\pi^*_{t,n}$ is an optimal solution for 
\begin{mini}
{\pi_t}{||\pi_t||_1}{}{}
\addConstraint{\Sigma^T_t\pi_t=\eta^*_t }{}{}
\label{chp6_linear_pro_proof}
\end{mini}
According to principle 4.5 in \cite{rardin1998optimization}, problem \eqref{chp6_linear_pro_proof} is equivalent to 
\begin{mini}
{\delta_t}{\mathbbm{1}^T\delta_t}{}{}
\addConstraint{\hat{\Sigma}^T_t\delta_t=\eta^*_t }
\addConstraint{\delta_t\geq0}
\label{chp6_linear_pro1}
\end{mini}
where $\delta_t=[\alpha_t^{(1)}, \alpha_t^{(2)},...,\alpha_t^{(n)},\beta_t^{(1)}, \beta_t^{(2)},...,\beta_t^{(n)}]^T$ satisfies $\alpha_t^{(i)}=\frac{|\pi_t^{(i)}|+\pi_t^{(i)}}{2}$, and $\beta_t^{(i)}=\frac{|\pi_t^{(i)}|-\pi_t^{(i)}}{2}$, with
\begin{equation}
\hat{\Sigma}_t=\left[\begin{array}{c}
\Sigma_t  \\
-\Sigma_t
\end{array}\right]=\left[\begin{array}{cc}
f^{11}_t & f^{12}_t\\
...&...\\
f^{n1}_t & f^{n2}_t\\
-f^{11}_t & -f^{12}_t \\
...&...\\
-f^{n1}_t & -f^{n2}_t 
\end{array}\right].
\end{equation}
Theorems 2.3 and 2.4 in \cite{bertsimas1997introduction} lists the necessary and sufficient conditions for the extreme point $\delta_t$, i.e.
\begin{enumerate}
    \item $\delta_t=[\delta_t^{(1)}, \delta_t^{(2)},...,\delta_t^{(n)},\delta_t^{(n+1)}, \delta_t^{(n+2)},...,\delta_t^{(2n)}]^T$.
    \item the $\hat{q}^{th}$ and $\hat{p}^{th}$ rows in $\hat{\Sigma_t}$  are linear independent,  $\delta_t^{(i)}=0$ if $i\neq \hat{q}$ or $\hat{p}$.
    \item $\delta_t$ is feasible solution.
\end{enumerate}
Without loss of generality, we assume the $p^{th}$ and $q^{th}$ rows in $\Sigma$ are linear independent, and we consider 4 cases:
\begin{equation}
\begin{split}
\delta_t^{[1]}&=\begin{cases}
[\delta_t^{[1],(1)}, \delta_t^{[1],(2)},...,\delta_t^{[1],(n)},\delta_t^{[1],(n+1)}, \delta_t^{[1],(n+2)},...,\delta_t^{[1],(2n)}]^T\\
\delta_t^{[1],(i)}=0 \quad \textit{ if $i\neq q$ or $p$}
\end{cases}\\
\delta_t^{[2]}&=\begin{cases}
[\delta_t^{[2],(1)}, \delta_t^{[2],(2)},...,\delta_t^{[2],(n)},\delta_t^{[2],(n+1)}, \delta_t^{[2],(n+2)},...,\delta_t^{[2],(2n)}]^T\\
\delta_t^{[2],(i)}=0 \quad \textit{ if $i\neq q+n$ or $p$}
\end{cases}\\
\delta_t^{[3]}&=\begin{cases}
[\delta_t^{[3],(1)}, \delta_t^{[3],(2)},...,\delta_t^{[3],(n)},\delta_t^{[3],(n+1)}, \delta_t^{[3],(n+2)},...,\delta_t^{[3],(2n)}]^T\\
\delta_t^{[3],(i)}=0 \quad \textit{ if $i\neq q$ or $p+n$}
\end{cases}\\\delta_t^{[4]}&=\begin{cases}
[\delta_t^{[4],(1)}, \delta_t^{[4],(2)},...,\delta_t^{[4],(n)},\delta_t^{[4],(n+1)}, \delta_t^{[4],(n+2)},...,\delta_t^{[4],(2n)}]^T\\
\delta_t^{[4],(i)}=0 \quad \textit{ if $i\neq q+n$ or $p+n$}
\end{cases}
\end{split}
\end{equation}
It is clear that there is a non-negative strategy in $\delta_t^{[1]}$, $\delta_t^{[2]}$, $\delta_t^{[3]}$ and $\delta_t^{[4]}$ because the $i^{th}$ row in $\hat{\Sigma}$ is the opposite of the $(i+n)^{th}$ row, and the non-negative strategy is feasible and an extreme point. This proves the existence of an extreme point for problem \eqref{chp6_linear_pro1}. Now, theorem 2.7 in \cite{bertsimas1997introduction} guarantees that there is an optimal solution which is an extreme point for problem \eqref{chp6_linear_pro1}.\\ 
With the second necessary and sufficient conditions of the extreme point, we know that an optimal solution $\delta_t^*$ for problem \eqref{chp6_linear_pro1} has at most two non-zero elements. This would imply an optimal solution, denoted by $\pi^*_{t,n}=[\pi_{t,n}^{(1)}, \pi_{t,n}^{(2)},...,\pi_{t,n}^{(n)}]^T$, for problem \eqref{chp6_linear_pro_proof} with at most two non-zero elements, which would also be the optimal strategy for \eqref{chp6_linear_pron}.\\ 
Without loss of generality, we assume $\pi_{t,n}^{(i)}=0$, $i\neq x,y$. $O_{t,2}=[O^{(x)}_t, O^{(y)}_t]$ and $\pi^*_{t,2}=[\pi_{t,n}^{(x)}, \pi_{t,n}^{(y)}]^T$ is a feasible strategy for problem \eqref{chp6_linear_pron} with $n=2$. We show that it is an optimal pair by contradiction.\\ 
If there is a feasible solution $\hat{O}_{t,n}=[\hat{O}^{(1)}_t, \hat{O}^{(2)}_t]$ and $\hat{\pi}^*_{t,2}=[\hat{\pi}_{t,2}^{(1)}, \hat{\pi}_{t,2}^{(2)}]^T$ such that $||\hat{\pi}^*_{t,2}||_1<||\pi^*_{t,2}||_1$, then $\hat{\pi}^*_{t,n}=[\hat{\pi}_{t,2}^{(1)}, \hat{\pi}_{t,2}^{(2)},0,...,0]^T$ is a feasible strategy for \eqref{chp6_linear_pron} such that $||\hat{\pi}^*_{t,n}||_1<||\pi^*_{t,n}||_1$, which is contradiction to our previous conclusion. Note that $||\pi^*_{t,2}||_1=||\pi^*_{t,n}||_1$, so problem \eqref{chp6_linear_pron} with $n=2$ and with $n\geq 2$ have the same minimum $\ell_1$ norm of allocation.

\subsection{Proof of Proposition \ref{chp6_propos_opt}}   
\label{chp6_appendix_opt}

According to the Bellman equation, the value function can be rewritten as,
\begin{equation}\label{chp6_eq:Voneperiod}
\begin{split}
V(t,W,\ln{S},H)&=\mathbb{E}_t(V(t+dt,W_{t+dt},H_{t+dt},\ln{S_{t+dt}})\mid W,H,\ln{S})\\
&=\max\limits_{\eta_{t}}\mathbb{E}_t(V(t+dt,W_{t+dt},H_{t+dt},\ln{S_{t+dt}})\mid W,\eta,H,\ln{S}).
\end{split}
\end{equation}
We expand  $V(t+dt,W_{t+dt},H_{t+dt},\ln{S_{t+dt}})$ at $t+dt$ in terms of all the variables.

\begin{equation}
\begin{split}
&V(t+dt,W_{t+dt},H_{t+dt},\ln{S_{t+dt}})=V(t+dt,W_{t},\ln{S_{t}},H_{t})+V_{W_t}(t+dt,W_{t},H_{t},\ln{S_{t}})d\hat{W}_t\\
&+\frac{1}{2}V_{W_tW_t}(t+dt,W_{t},H_{t},\ln{S_t})(d\hat{W}_t)^2+ V_{\ln{S_t}}(t+dt,W_{t},H_{t},\ln{S_t})d\ln{S_t} 
+  V_{H_t}(t+dt,W_{t},H_{t},\ln{S_t})dH_t \\
&+\frac{1}{2}  V_{\ln{S_t}\ln{S_t}}(t+dt,W_{t},H_{t},\ln{S_t})d\ln{S_t}d\ln{S_t} +\frac{1}{2}  V_{H_tH_t}(t+dt,W_{t},\ln{S_t},H_{t})dH_tdH_t \\
&+V_{W_t\ln{S_t}}(t+dt,W_{t},H_{t},\ln{S_t})d\hat{W}_td\ln{S_t} +   V_{W_t H_t}(t+dt,W_{t},H_{t},\ln{S_t})d\hat{W}_tdH_t\\
&+  V_{\ln{S_t}H_t}(t+dt,W_{t},\ln{S_t},H_{t})d\ln{S_t}dH_t +o(dt).
\end{split}\label{chp6_HJB_lsmc1}
\end{equation}

Substituting $d\hat{W}_t$, $d\ln{S_t}$, $dH_t$ which can be found in equation \eqref{chp6_gene_diff_model}, taking conditional expectation on both sides, and rewriting $V(t,W_{t},H_{t},\ln{S_{t}})$ in a quadratic form with respect to $\eta$ leads to

\begin{equation} 
\begin{split}
V(t,W_{t},H_{t},\ln{S_{t}})&=\max_{\eta_t} \left( \sum_{i,j=1}^2 f_{i,j}(t,W_t,,H_t\ln{S_{t}})\eta^{(i)}_t \eta^{(j)}_t+ \sum_{i=1}^2 f_i(t,W_t,H_t,\ln{S_{t}})\eta^{(i)}_t+f_0(t,W_t,H_t,\ln{S_{t}})\right)\\
f_{i,j}(t,W_t,H_t,\ln{S_{t}})&=\frac{1}{2} V_{W_tW_t}(t+dt,W_{t},H_{t},\ln{S_{t}})\hat{W}_t^2 (\Phi\Phi^T)_{i,j}  dt\\
f_{i}(t,W_t,H_t,\ln{S_{t}})&=V_{W_t}(t+dt,W_{t},H_{t},\ln{S_{t}})\hat{W}_t \Lambda_i dt+  V_{W_t\ln{S_t}}(t+dt,W_{t},H_{t},\ln{S_t})\hat{W}_t \sigma_S B_i dt \\
& + V_{W_tH_t}(t+dt,W_{t},H_{t},\ln{S_t})\hat{W}_t \sigma_H A_i dt\\
f_0(t,W_t,H_t,\ln{S_t})&=V(t+dt,W_{t},H_{t},\ln{S_t})
+V_{W_t}(t+dt,W_{t},H_{t},\ln{S_t})\hat{W}_t r dt\\
&+V_{\ln{S_t}}(t+dt,W_{t},H_{t},\ln{S_t})(\lambda^S\sigma^S-\frac{1}{2}(\sigma^S)^2) dt+  V_{H_t}(t+dt,W_{t},H_{t},\ln{S_t})\mu^H dt\\
&+  \frac{1}{2}V_{H_tH_t}(t+dt,W_{t},H_{t},\ln{S_t}) (\sigma^H)^2 dt+\frac{1}{2}  V_{\ln{S_t}\ln{S_t}}(t+dt,W_{t},H_{t},\ln{S_t}) (\sigma^S)^2 dt\\
&+  V_{\ln{S_t}H_t}(t+dt,W_{t},H_{t},\ln{S_t}) \sigma^S\sigma^H dt.
\end{split}\label{chp6_HJB_lsmc}
\end{equation}

We assume a sufficiently small $dt$ so that $o(dt)$ terms are omitted when taking conditional expectations. The optimal allocation is given by the solution to the system of equations:

\begin{equation}
\sum_{j=1}^2 2f_{i,j}(t,W_t,H_t,\ln{S_{t}}) \eta^{(*,j)}_t = -  f_{i}(t,W_t,H_t,\ln{S_{t}}), \: i=1,2.
\label{chp6_pi_opt_system}
\end{equation}

With the representation of the value function in equation \eqref{chp6_valuefunc_ansatz} and assuming that $f(t,H,\ln{S})=exp(P_k(t,H,\ln{S}))$, the derivatives of value function with respect to each stock and state variable can be rewritten as,

\begin{equation}
\begin{split}
V_{W}(t+dt,W_t,H_t,\ln{S_t})&=\hat{W}_t^{-\gamma}exp(P_k(t+dt,H_t,\ln{S_t}))\\
V_{WW}(t+dt,W_t,H_t,\ln{S_t})&=-\gamma \hat{W}_t^{-\gamma-1}exp(P_k(t+dt,H_t,\ln{S_t}))\\
V_{W\ln{S_t}}(t+dt,W_t,H_t,\ln{S_t})&=\hat{W}_t^{-\gamma}exp(P_k(t+dt,H_t,\ln{S_t}))\frac{\partial P_k(t+dt,H_t,\ln{S_t})}{\partial \ln{S_t}}\\
V_{WH_t}(t+dt,W,H_t,\ln{S_t})&=\hat{W}_t^{-\gamma}exp(P_k(t+dt,H_t,\ln{S_t}))\frac{\partial P_k(t+dt,H_t,\ln{S_t})}{\partial H_t}.
\end{split}\label{chp6_deri_v}
\end{equation}
Substituting \eqref{chp6_deri_v} into \eqref{chp6_pi_opt_system}, the optimal strategy can be approximated as follows:

\begin{equation}
\begin{split}
\sum_{j=1}^2 g_{i,j}(t,W_t,H_t,\ln{S_{t}}) \eta^{(*,j)}_t &=  g_{i}(t,W_t,H_t,\ln{S_{t}}), \: i=1,2\\
g_{i,j}(t,W_t,H_t,\ln{S_{t}})&= \gamma  (\Phi\Phi^T)_{i,j}  \\
g_{i}(t,W_t,H_t,\ln{S_{t}})&= \Lambda_i+  \frac{\partial P_k(t+dt,H_t,\ln{S_t})}{\partial \ln{S_t}} \sigma^S B_i+\frac{\partial P_k(t+dt,H_t,\ln{S_t})}{\partial H_t}\sigma^S A_i, \\
\end{split}
\end{equation}
Then, the optimal strategy can be rewritten in matrix form:
\begin{equation}
    \eta_t^*=\frac{1}{\gamma}(\Phi \Phi ^{T})^{-1} (\Lambda + \frac{\partial P_k}{\partial H}\sigma^HA+\frac{\partial P_k}{\partial \ln{S}}\sigma^SB ).
\end{equation}
\section{Alternative approximation method and comparison}
\label{chp6_direct}
\subsection{Direct method}

We introduced an alternative method for derivatives-based portfolio strategy, namely PAMC-direct method, which is straightforward application of the PAMC. At each re-balancing time, the path-wise  option price $O_t$,  Delta $\frac{\partial O_t}{\partial S_t}$ and the sensitivity to the state variable $\frac{\partial O_t}{\partial H_t}$ are approximated, so the instantaneous dynamics of derivatives are obtained. In this way, derivatives can be taken as an asset with dynamics known explicitly, the PAMC method is directly applied.  Next proposition shows the estimation of optimal strategy $\pi_t^*$ in PAMC-direct.
\begin{proposition}
Given the approximation of the value function at the next re-balancing time $t+\Delta t$  (i.e. $\frac{W^{1-\gamma}}{1-\gamma} \exp\{P_k\}(t+\Delta t,H,\ln{S})$), the optimal strategy at time $t$ is given by
\begin{equation}
  \pi_t^*=\frac{1}{\gamma}(\Sigma _{t}\Phi \Phi ^{T}\Sigma_{t}^{T})^{-1} (\Sigma _{t}\Lambda + \frac{\partial P_k}{\partial H}\sigma^H\Sigma _{t}A+\frac{\partial P_k}{\partial \ln{S}}\sigma^S\Sigma _{t}B ).
  \label{chp6_opt_pi_direct}
\end{equation}

\begin{proof}
Similar to Appendix \ref{chp6_appendix_opt}.

\end{proof}
\end{proposition}

We continue to use the notation in Table \ref{chp6_note} and describe the step by step algorithm of the PAMC-direct in Algorithm \ref{chp6_PAMH_deri1}.

\begin{algorithm}[h]
\caption{PAMC-direct method} \label{chp6_PAMH_deri1}
      \KwIn{$S_0$,$W_0$,$H_0$\\}
      \KwOut{Optimal trading strategy $\pi^*_0$}
      initialization\;
      Generating $n_r$ paths of $B_t^m$, $B_t^{m,H}$,$S_t^{m}$, $H_t^m$ $ \quad for \quad m= 1 ...n_r$\;
      Apply approximation methods and obtain the price of  $O_t(H_t^m,\ln{S_t^{m}})$ as well as its sensitivity $\frac{\partial O_t}{\partial S_t}(H_t^m,\ln{S_t^{m}})$ and $\frac{\partial O_t}{\partial H_t}(H_t^m,\ln{S_t^{m}})$ $for$ $t=0,\Delta t, ...T$ \;
      \While{ $t=T-\Delta t$ }{
      \For{$m= 1 ...n_r$}{
        Compute the variance matrix $\Sigma_{T-\Delta t}^m$ with derivatives price and sensitivity obtained in step 3 \;
        Directly compute optimal allocation $\pi^m_{T-\Delta t}$ with Equation \eqref{chp6_opt_pi_direct} where the $P_k=1$ at time $T$\;
        \For{ $n= 1 ...N$}{
        Generate $\hat{S}_T^{m,n}$ and $\hat{H}_T^{m,n}$ given $S_{T-\Delta}^{m}$ and $H_{T-\Delta}^{m}$ and obtain $\hat{O}_T^{m,n}$\;}
        Compute wealth $\hat{W}_T^{m,n}(\pi^m_{T-\Delta t})$ at the terminal given the wealth at $W_{T-\Delta t}=W_0$, the transformed value function is estimated by
        $\hat{v}^{m}=\ln{[(1-\gamma)\frac{1}{N}\sum\limits_{n=1}^{N}U(\hat{W}_T^{m,n}(\pi^m_{T-\Delta t}))]}-(1-\gamma)\ln{W_0}$  \;}
        Regress $\hat{v}^{m}$ over the polynomial of $H_{T-\Delta t}^m$ and $\ln{S}_{T-\Delta t}^m$, and obtain the function $L_{T-\Delta t}(H,\ln{S})$\; }

       \For{$t=T-2\Delta t$ to $\Delta t$}
        {      \For{$m= 1 ...n_r$}{
        Compute the variance matrix $\Sigma_{t}^m$ with derivatives price and sensitivity obtained in step 3\;
        Directly compute optimal allocation $\pi^m_{t}$ with Equation \eqref{chp6_opt_pi_direct} where the $P_k=L_{t+\Delta t}(H,\ln{S})$\;
        \For{ $n= 1 ...N$}{
        Generate $\hat{S}_{t+\Delta t}^{m,n}$ and $\hat{H}_{t+\Delta t}^{m,n}$ given $S_{t}^{m}$ and $H_{t}^{m}$ and obtain $\hat{O}_{t+\Delta t}^{m,n}$\;}
        Compute wealth $\hat{W}_{t+\Delta t}^{m,n}(\pi^m_{t})$ at the terminal given the wealth at $W_t=W_0$, the transformed value function is estimated by
         $\hat{v}^{m}=\ln{[\frac{1}{N}\sum\limits_{n=1}^{N}(W^{m,n}_{t+\Delta t}(\pi^m_t))^{1-\gamma}exp(L_{t+\Delta t}(\hat{H}_{t+\Delta t}^{m,n},\ln{\hat{S}}_{t+\Delta t}^{m,n}))]} -(1-\gamma)\ln{W_0}$\;}
        Regress $\hat{v}^{m}$ over the polynomial of $H_{t}^m$ and $\ln{S}_{t}^m$, and obtain the function $L_{t}(H,\ln{S})$\; }
        
        \While{ $t=0$ }{ Compute the variance matrix $\Sigma_0$ with derivatives price and sensitivity obtained in step 3, and the optimal allocation $\pi_0^*$ is obtained with Equation \eqref{chp6_opt_pi_direct} and where the $P_k=L_{\Delta t}(H,\ln{S})$\;
       
        }

         return $\pi^*_{0}$
\end{algorithm}

\subsection{Comparison between the PAMC-direct method and the PAMC-indirect method}
\label{chp6_heston_com}

In this section, we implement the PAMC-direct method and the PAMC-indirect method on the Heston SV model given in \eqref{chp6_heston} for comparison purpose.   The derivatives-based portfolio given the Heston model was first studied in \cite{liu2003dynamic}, where the author constructed a portfolio with derivative securities and a stock so that volatility risk is able to be managed.  The optimal strategy stock-derivatives portfolio is solved in closed-form. The accuracy and efficiency of the PAMC-direct and the PAMC-indirect are examined in comparison with the analytical solution.

We continue to use the market-calibrated set of parameters in Table \ref{chp6_para_heston}. For simplicity,  we let $O_t$ be a delta-neutral straddle because the delta-neutral position keeps the straddle near-the-money, and the liquidity should not be a concern.

\begin{figure}[H]
	\centering
	\vspace{0.35cm}
	\subfigtopskip=2pt
	\subfigbottomskip=2pt
	\subfigcapskip=-5pt

	\subfigure[allocation on stock]{
		\includegraphics[width=0.32\linewidth]{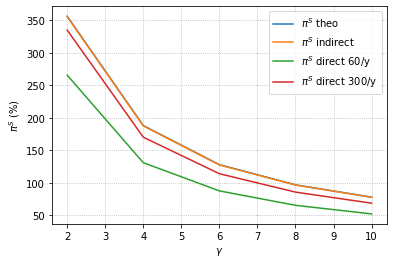}}
		\subfigure[allocation on straddle option]{
		\includegraphics[width=0.32\linewidth]{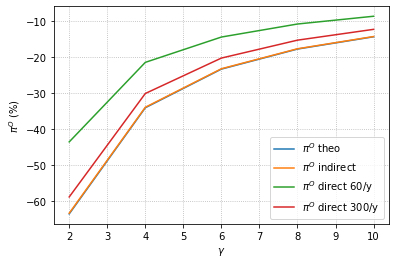}}
		\subfigure[Computational time]{
		\includegraphics[width=0.32\linewidth]{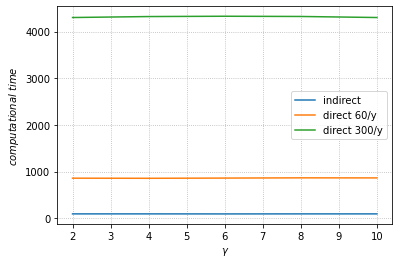}}

	\caption{Allocation on straddle versus $\gamma$}
    \label{chp6_com_di_in}
\end{figure}

Figure \ref{chp6_com_di_in} (a) and (b) compares the optimal allocation on the stock and straddle across different values of risk aversion level $\gamma$. We let the re-balancing frequency of the PAMC-indirect method  be $60$ times per year, i.e. investors roughly adjust their positions weekly. Optimal allocation from the PAMC-indirect method and theoretical solution (re-balancing continuously) are visually overlapped, the PAMC-indirect method exhibits very excellent accuracy in this case. The allocation from the PAMC-direct method with $60$ re-balances per year is subject to a substantial error, on the other hand, the gap to the theoretical solution shrinks if we let the re-balancing frequency be $300$ times per year (roughly daily re-balance). We expect the gap will vanish as re-balancing frequency continues to increase. The computational times of the PAMC-direct and PAMC-indirect methods are compared in figure \ref{chp6_com_di_in} (c), the time required for the PAMC-indirect method is significantly smaller than the time for the PAMC-direct method. The PAMC-indirect is superior to the PAMC-direct with regard to both accuracy and computational efficiency, we hence use only the PAMC-indirect in section \ref{chp_6_sec4}.

\end{document}